\documentstyle[12pt,epsf,epsfig,wrapfig]{article}
\def\be{\begin{equation}}
\def\ee{\end{equation}}
\newcommand{\bea}{\begin{eqnarray}}
\newcommand{\eea}{\end{eqnarray}}

\begin{document}

\begin{center}
{\bfseries PRODUCTION OF SPIN-3 MESONS IN DIFFRACTIVE DIS}

\vskip 5mm
I.P. Ivanov

\vskip 5mm
{\small
{\it
University of Liege, Belgium
}\\
{\it
Sobolev Institute of Mathematics, Novosibirsk, Russia
}\\
$\dag$ {\it
E-mail: Igor.Ivanov@ulg.ac.be
}}
\end{center}

\vskip 5mm
\begin{abstract}
Non-trivial spin effects do not die out with energy growth. Here, we discuss one
example of such effects --- production of spin-3 mesons in diffractive DIS.
Using both explicit $k_t$-factorization calculations and their
vector-dominance-model interpretation, we argue that diffractive
production of $\rho_3(1690)$ is a unique probe of several novel aspects of diffraction.
\end{abstract}

\vskip 8mm

It is well known that even at highest energies a significant fraction
of hadron-hadron collisions must be elastic. Elastic scattering is a member
of the family of diffractive processes, in which the colliding hadrons
can survive the scattering or turn into a small-mass diffractive system, $M_{diff} \ll\sqrt{s}$.
In deep-inelastic scattering, DIS, where the virtual photon can be also viewed as a hadron,
a significant part of all $\gamma^* p$ collisions is also diffractive.

The $t$-channel exchange that drives diffraction, the Pomeron, is often pictured
as a ``spin-blind'' object. This leads to a prejudice that all non-trivial spin effects
must die out at high energies, where the Pomeron exchange dominates over the secondary Reggeons.
Partly in order to eliminate this prejudice, we present here our recent results
on production of spin-3 mesons in diffractive DIS, \cite{spin3,VDMlimits},
which is a genuine example of {\em non-trivial spin effects in diffraction}.

\section{Basics of diffractive meson production}

Dynamics of diffractive DIS is conveniently described within the color dipole
approach, \cite{colordipole}. The incoming photon turns into a $q\bar q$ pair (a color dipole),
which experiences scattering off the target and then is projected onto the final meson $V$.
Thanks to the Lorentz-dilatation of the transverse motion inside the projectile,
the amplitude for this transition can be written in the probabilistic form
$$
{1 \over s}{\cal A}(\gamma \to V) = \langle V| \hat \sigma|\gamma \rangle =
\int dz\, d^2\vec r\, \Psi_V^*(z,\vec r) \sigma_{\mathrm{dip}}(\vec r) \Psi_\gamma(z,\vec r)\,.
$$
Here, $\hat \sigma_{\mathrm{dip}}$ is the {\em diffraction operator}, which acts in the projectile Fock space.
In the $\vec r$-space it is diagonal and is known as {\em dipole cross section}.

Practical calculations are most convenient in the transverse momentum space.
In this approach, known as the $k_t$-factorization approach, the amplitude has form:
\be
{1 \over s}{\cal A}(\gamma \to V) =  {e\, c_V \over 4 \pi^2}
\int {dz\, d^2\vec{k}_\perp\over z(1-z)}  \int {d^2\vec \kappa \over \vec \kappa^4}
\alpha_s\,{\cal F}\cdot
I^{V}_{\lambda_V;\lambda_\gamma}\cdot\Psi_V(p^2)\,.\label{amp}
\ee
Here $z$ is the lightcone momentum fraction of the photon carried by the quark,
$\vec k$ is the relative transverse momentum of the $q\bar q$ pair, while
$\vec \kappa$ is the transverse momentum of the gluon.
Coefficient $c_V$ is the standard flavor-dependent average charge
of the quark.
The color dipole cross section is encoded via the unintegrated gluon distribution
function ${\cal F}$. In our calculations we used fits of ${\cal F}$
obtained in \cite{in2000} and adapted to the off-forward kinematics needed for the meson production, see details
in \cite{review}.

This approach can be used to calculate production of quarkonia in different spin-orbital states.
The only requirement is that $P=C=-1$.
\begin{itemize}
\item Ground state vector mesons ($L=0$, $n_r=0$): $\rho, \omega, \phi, J/\psi, \Upsilon$.
\item Radially excited VM ($L=0$, $n_r>0$): $\approx \rho'(1450)$, \dots
\item Orbitally excited VM ($L=2$, $n_r=0$): $\approx \rho''(1700)$, \dots
\item High-spin mesons, e.g. spin-3 mesons with $L=2$ such as $\rho_3(1690)$.
\end{itemize}
The properties of the given meson appear in (\ref{amp}) in two ways:
via the radial wave function $\Psi_V(p^2)$, and via the spin-orbital structure,
which is encoded in an appropriate spinorial structure of the $q\bar q V$ vertex, \cite{in99},
and is present in (\ref{amp}) implicitly inside the integrands $I^{V}_{\lambda_V;\lambda_\gamma}$.
For example, for the $D$-wave vector meson the vertex has form
$\bar u \Gamma_D^\mu u \cdot V_\mu$, where $V_\mu$ is the polarization vector and
$\Gamma^\mu_D = \gamma^{\mu} - 4(M+m)p^\mu/(M^2-4m^2)$.
The corresponding structure for the spin-3 meson is $u \Gamma^{\mu\nu\rho} u\cdot T_{\mu\nu\rho}$,
where $T_{\mu\nu\rho}$ is the polarization tensor and $\Gamma^{\mu\nu\rho}$ was derived in \cite{spin3}.

\section{Production of orbitally or spin-excited mesons}

\subsection{Characteristic features}
Let us first discuss the vital property of diffraction:
{\em it does not conserve orbital momentum $L$ of the $q\bar q$ pair}.
Indeed, symbolically the amplitude is
\be
{\cal A} \propto  \int {dz d^2\vec{k}_\perp \over z(1-z)} \,
\langle L'| \hat\sigma_{\mathrm{dip}} | L \rangle
= \int {4 \over M} d^3 p\,
\langle L'| \hat\sigma_{\mathrm{dip}} | L \rangle
\not = 0\,,\label{nondiagonal}
\ee
because diffraction operator $\hat\sigma_{\mathrm{dip}}$ is not spherically symmetric.
Namely, when calculating the diagrams, one observes that the transverse momentum circulating
in the quark loop is much more important than the longitudinal one.
This is the unavoidable consequence of the fact that any collision has a preferred
direction.

Since the incoming photon can be also represented as a coherent combination of
various vector mesons, including radially and orbitally excited mesons, see discussion
in \cite{VDMlimits}, one can, therefore, expect that
orbitally excited and spin-excited mesons will be present among the final states in diffractive DIS.

Orthogonality of $q\bar q$ with different $L$ suppresses the helicity conserving,
but not the helicity violating amplitudes. Therefore, a much stronger
helicity violation is expected for orbital excitations than for
grounds state mesons.

\subsection{Numerical results}

Numerical results are obtained by direct evaluation of amplitudes (\ref{amp}) and integration
of the differential cross section within the diffraction cone, $|t|<1$ GeV$^2$.
We took into account all helicity amplitudes, including all helicity violating transitions.

\begin{wrapfigure}[15]{R}{7.4cm}
\vspace{-1cm}
\begin{center}
\mbox{\epsfig{figure=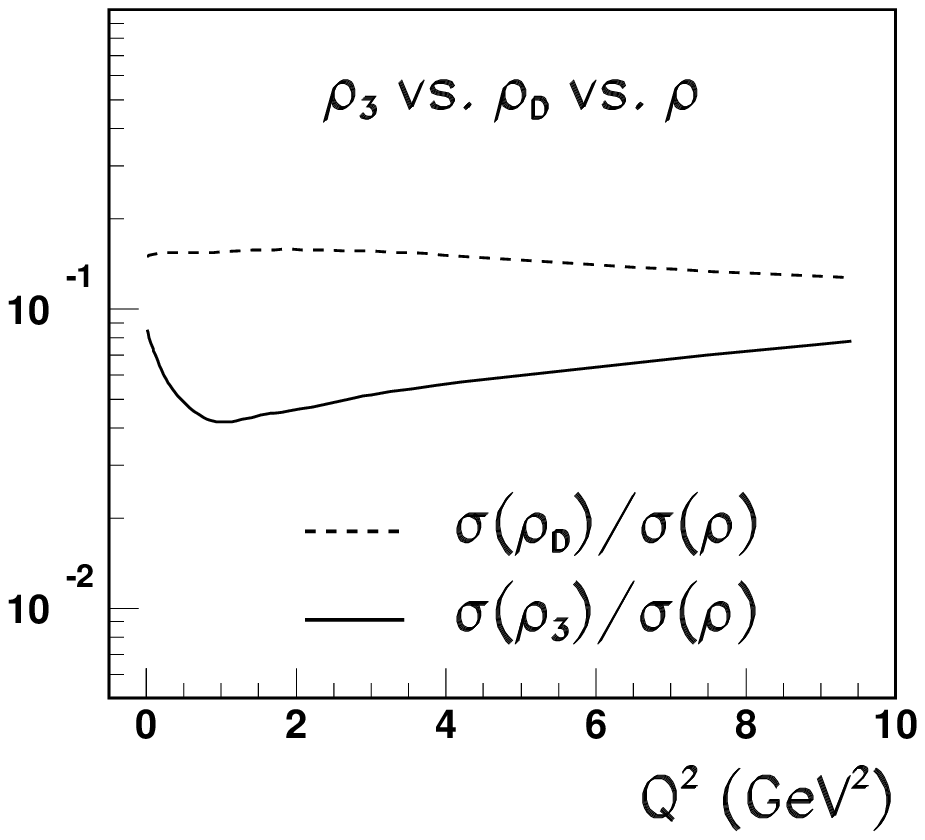,width=7cm}}
\end{center}
{\small{\bf Figure 1.} Ratios $\sigma(\rho''(1700))/\sigma(\rho)$
and $\sigma(\rho_3(1690))/\sigma(\rho)$ as functions of $Q^2$.}
\label{IvanovI_fig1}
\end{wrapfigure}

The main sources of uncertainty in the absolute values of the cross sections are the
choice of the parametrization of the radial wave function and the $e^+e^-$ decay width, which is used
to adjust the size of the wave function. Our experience with ground state VM production \cite{review}
shows that different choices of the wave function lead to results
differing by factor of 2. The results for the orbitally excited mesons are even more uncertain
since the value of $\Gamma(\rho''(1700)\to e^+e^-)$ is known only within factor of 5.
However, the relative production rates of $\rho''(1700)$ and $\rho_3(1690)$ are more stable,
within factor of 2.

Fig.~1 shows the results for the ratios $\sigma(\rho''(1700))/\sigma(\rho)$ and
$\sigma(\rho_3(1690))/\sigma(\rho)$ as functions of $Q^2$. Both ratios are ${\cal O}(0.1)$
and are comparable. Since $\rho''(1700)$ and $\rho_3(1690)$ are degenerate,
this means that one must perform a very careful analysis of multipion diffractive final states
in order to separate these two mesons.

Studying $\rho_3$ production in more detail, we found some other peculiar features.
\begin{itemize}
\item
Numerical calculations confirm very large contribution from helicity violating
transitions in $\rho_3$ even at moderate $Q^2$. We even predict {\em domination of
helicity violation} at small $Q^2$ --- a new regime in diffraction.
\item
The radial wave functions of the orbitally excited mesons are broader than of the ground
states. Thus, typical dipole sizes in $\rho_3$ photoproduction are $\sim 1.5$ times larger (up to 2 fm)
than for $\rho$ photoproduction.
\item
$\sigma_L/\sigma_T$ ratio is abnormally large for $\rho_3$.
\end{itemize}

\subsection{Coupled channel analysis}

The cross sections for $\rho''(1700)$ and $\rho_3(1690)$ production are of the same orders of magnitude,
yet, the mechanisms of their diffractive production are quite different.
We found this by performing, in the spirit of generalized vector dominance model,
a coupled channel analysis of the action of diffraction operator
in the Fock space generated by three mesons: ground state $\rho$, orbitally excited state $\rho''(1700)$
and spin-3 meson $\rho_3(1690)$. Details are reported in \cite{VDMlimits}. Here we just show
the matrix for the integrated cross sections and with the sum over all final polarization states:
\be
\sigma_{ba} = \langle V_b|\hat\sigma_{\mathrm{dip}}| V_a\rangle = \left(
\begin{array}{ccc}
19 & 1 & 0.2 \\
1 & 27 & 0.3 \\
1.3 & 0.4 & 19
\end{array}\right)\ \mbox{mb}\,,\quad V_a, V_b = \rho,\rho'',\rho_3\,.
\nonumber
\ee
The diagonal elements have uncertainty of $\sim 50\%$, while the off-diagonal elements
are uncertain within factor 2-3.

Since the hadronic part of the incoming photon at small virtuality can be roughly represented as
$|\gamma\rangle_h \sim |\rho_S\rangle + 0.2 |\rho_D\rangle$, one can conclude that
\begin{itemize}
\item
$\rho''(1700)$ is produced mostly via ``direct materialization'' of the $D$-wave component
of the photon followed by diagonal scattering, $\gamma \to \rho'' \rightarrow \rho''$.
\item
$\rho_3(1690)$ is produced via truly off-diagonal transition $\gamma \to \rho \rightarrow \rho_3$.
\item
Thus, $\rho_D$ and $\rho_3$ probe {\em different properties of diffraction}.
\end{itemize}

\section{Experimental opportunities}

In contrast to the ground state vector meson production, \cite{review}, the data
on excited VM, in particular, on the orbitally or spin-excited mesons, are very scarce.
For example, the only published data on excited $\rho_3(1690)$ go back to 1986, when the fixed-target experiment
OMEGA at CERN measured it in diffractive multipion photoproduction, \cite{omega1986}.

Modern era experiments, both collider or fixed target, have great potential in making much progress
in this field. What one needs is to study diffractive multipion production and extract the resonant
contribution. One broad peak at $M\approx 1.1-1.9$ GeV should be separated into three excited $\rho$ states:
$\rho'(1450)$, $\rho''(1700)$, and $\rho_3(1690)$.

Specifically, the best tool to extract the $\rho_3$ contribution would be the partial-wave analysis.
We expect that the most sensitive to the $\rho_3$ would be $\pi^+\pi^-$ final state
at not too small $t$, say, at $|t|\approx 0.5$ GeV$^2$. One can try to see the $\rho_3$ by comparing
multipion spectra in diffractive production and in $e^+e^-$ annihilation experiments, since
$\rho_3$ is present in diffraction but absent in the annihilation. Preliminary analysis \cite{VDMlimits}
gave interesting results.
\\

{\bf Acknowledgements}. The work was supported by FNRS
and partly by grants RFBR 05-02-16211 and NSh-5362.2006.2.

\end{document}